# Towards Building A Facial Identification System Using Quantum Machine Learning Techniques

Philip Easom-McCaldin, Ahmed Bouridane, Ammar Belatreche
Dept. of Computer and Information Sciences, Northumbria University, Newcastle upon Tyne, UK
Email: {philip.easom,ahmed.bouridane,ammar.belatreche}@northumbria.ac.uk

Richard Jiang
Dept. of Computing and Communications, Lancaster University, Lancaster, UK
Email: r.jiang2@lancaster.ac.uk

*Abstract*— **In the modern world, facial identification is an extremely important task in which many applications rely on high performing algorithms to detect faces efficiently. Whilst classical methods of SVM and k-NN commonly used may perform to a good standard, they are often highly complex and take substantial computing power to run effectively. With the rise of quantum computing boasting large speedups without sacrificing large amounts of much needed performance, we aim to explore the benefits that quantum machine learning techniques can bring when specifically targeted towards facial identification applications. In the following work, we explore a quantum scheme which uses fidelity estimations of feature vectors in order to determine the classification result. Here, we are able to achieve exponential speedups by utilizing the principles of quantum computing without sacrificing large proportions of performance in terms of classification accuracy. We also propose limitations of the work and where some future efforts should be placed in order to produce robust quantum algorithms that can perform to the same standard as classical methods whilst utilizing the speedup performance gains.**

## I. Introduction

In recent years, quantum computing has become a promising area of research with the promise of many rich, performance enhancing benefits due to natural quantum behaviours and principles. One of the most promising of areas is quantum machine learning. With the rise of quantum machine learning being propelled towards benefits of performance speedups, quantum algorithms have already produced speedups of exponential factors over many classical algorithms currently used [1].

Facial identification within images is a highly critical task within machine learning that has many real-world applications. Within humans, facial identification within images in a simple task, but from a computer vision perspective, the task is highly complex. Nevertheless, by applying machine learning techniques, facial identification has been able to be applied to a wide variety of domains that assist us in day to day living.

Examples of applications are plentiful across the modern world, through biometric analysis, healthcare, security and marketing. Because of the high variety in applications, it is critical that we are able to innovate our approaches to tasks such as facial identification in order to overcome drawbacks and challenges we are currently faced with.

Currently, much focus within the NISQ era of quantum machine learning has been focused towards classification using variational circuits [2 - 8] as a step towards creating robust quantum neural networks [9, 10] and advanced deep-learning algorithms [11, 12]. Efforts have also been conducted towards quantum k-NN algorithms that can conduct efficient searches of data stored in superpositions using subroutines [13, 14]. However, there have not been many works that showcase quantum algorithms applied in practice to real-world tasks [15, 16, 17]. Many algorithms proposed have relied on synthetic datasets [18, 19] or are purely theoretical [20 - 24] and we will not be able to validate these systems until improvements have been made to current quantum hardware.

In this paper, we aim to explore how we can use quantum machine learning techniques to build classifiers based on real-world applications, specifically towards facial identification within this work. Here, we look to build a classifier using foundational quantum subroutines that estimate fidelities between quantum states. This gives an initial insight into how quantum algorithms may perform at a simple level, and gain perspective on where efforts may be placed to improve upon these initial results to create robust classification systems that have potential to be applied to real-world tasks in the future.

The organisation of the paper is as follows. Firstly, a brief overview of quantum computing is given to give some slight background knowledge. Then we describe the systems design and methodology used to perform facial identification. Afterwards, an analysis of the systems performance is given in comparison to common classical algorithms, in particular a support vector machine (SVM) and k-nearest neighbour (k-NN). Finally, a conclusion is drawn and future directions are considered.

## II. QUANTUM COMPUTING OVERVIEW

In classical computing, the basic unit of information is the bit. However, in quantum computing, this unit is known as the qubit. Qubits represent information in a 2-dimensional complex vector space, known as Hilbert space. This allows us to represent classical information very efficiently in feature spaces that are not tractable or feasible to do so in classical systems. In order to manipulate qubit states, we can use a collection of unitary matrices to transform the state of the qubit with respect to the matrix used. These matrices are represented in the form of quantum gates, where visualizing the quantum circuit that operates on an input state, we can see what transformations are occurring throughout.

### A. Quantum Encoding

In order to perform quantum computations using classical information, it must first be transformed into its' quantum form. Whilst many quantum representations of images and classical data have been presented [25-28], basis encoding and amplitude encoding are two common methods of doing this.

*Basis encoding:* Within basis encoding, inputs consisting of a binary string become represented by qubits encoded in the quantum computational basis states. For example, the binary string $\psi = 010$ becomes encoded as the quantum state $\psi = |010\rangle$. Using this method of encoding, we require 1 qubit per bit of input information.

*Amplitude Encoding:* A different approach to qubit encoding is through amplitude encoding. Here, a normalized input vector is taken and encoded onto the associated amplitudes of a quantum state. For example, a 2-dimensional vector $\begin{bmatrix}0.6\\0.8\end{bmatrix}$ can be represented as the quantum state $\psi = 0.6|0\rangle + 0.8|1\rangle$. In doing so, $n$ qubits can represent $2^n$ units of information. This presents a much more efficient representation of classical information than when compared to basis encoding.

## III. SYSTEM DESIGN

### A. System design overview

The design of the system can be split into three main sections. Firstly, input data is pre-processed classically. Secondly, features are encoded onto the quantum circuit and the routine is executed. Thirdly, circuit measurement occurs for post-processing and classification

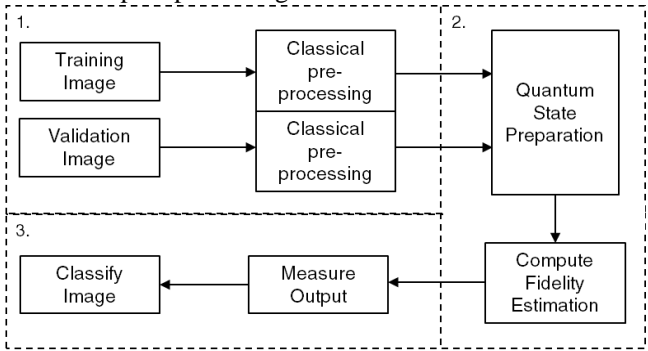

Figure 1. System overview flowchart

### B. Data pre-processing & feature encoding

For the purposes of this work, the AT&T face dataset was used. This dataset consists of 400 greyscale images of faces with varying angles and expressions. This dataset was consequently split into 300 images for training and the remaining 100 for validation. For the purposes of validation, a bespoke dataset consisting of 400 non-face objects and images was included. All input images were initially resized to a size of 8x8 and flattened to form a 64-dimensional vector. Once the data had been pre-processed, an average face vector was formed from the training images to create a state to compare against.

$$x = \frac{1}{N}\sum_{i=0}^{N} T_i$$

Where $N$ is the number of training images and $T$ is the image vector. In order to encode data onto the quantum circuit, amplitude encoding is used to create the input state vectors. For the input vector of $D = 64$ dimensions, this would require $\log_2(D) = 6$ qubits per input state. The input vector, $X$, has to be normalized to have a sum-of-squares equalling one.

$$X = \frac{x}{\|x\|} \quad \rightarrow \quad |\psi\rangle = \sum_{i=0}^{D} X_i |i\rangle$$

The input image state vectors can now be encoded as quantum states $|\psi\rangle$, being the comparison state and $|\phi\rangle$, being the test state. Once amplitude encoding has been performed on both of the input state vectors, the system, $S$, is now in the following state.

$$|S\rangle = |0\rangle \otimes |\psi\rangle \otimes |\phi\rangle$$

### C. Circuit design

After the input state vectors have been produced, we can initialize the quantum circuit before execution. The total number of qubits required, $(Q)$, is:

$$Q = 2\log_2(D) + 1$$

where $D$ is the dimensionality of the input vector and the additional 1 is for the ancilla qubit which holds our fidelity measurement. Here, the input test state and the comparison state are encoded leaving 1 ancilla qubit for comparing fidelities.

In order to differentiate fidelity between states, a subroutine known as a SWAP test is performed. The SWAP test estimates inner-products between two quantum states, i.e. $F = |\langle\psi|\phi\rangle|^2$ and is the most computationally expensive portion of the system. A SWAP test requires an ancilla qubit to contain the result and is split into three parts.

First, a Hadamard operation, H, is applied to the ancilla qubit initialized in the $|0\rangle$ state. This places the qubit into an equal superposition of the computational basis states.

$$|0\rangle \rightarrow \frac{|0\rangle + |1\rangle}{\sqrt{2}}$$

This is followed with a swap operation acting on the encoded quantum states. The third stage of the SWAP test routine is ended with another Hadamard operation acting on the ancilla qubit. Below shows the circuit diagram used for the purposes of this work, where $|\psi\rangle$ is the train state and $|\phi\rangle$ is the input test state being compared.

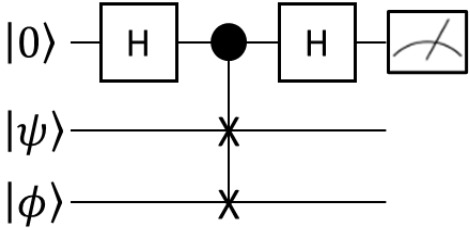

Figure 1. Circuit diagram for the quantum facial identification system

After the swap test has been applied, the total state of the system is as follows.

$$|S\rangle = \frac{1}{\sqrt{2}}(|0\rangle|\psi\rangle|\phi\rangle + |1\rangle|\psi\rangle|\phi\rangle).$$

#### D. Post-processing & classification

Once both of the quantum states have been encoded and a SWAP test has been performed, the output must be measured to be post-processed and produce a classification result. Within the system, only the ancilla qubit used within the SWAP test is measured with respect to the Z-basis in order to produce the fidelity of the two encoded states. After the second Hadamard operation has been performed, the measurement output of the ancilla qubit is as follows.

$$P(0) = \frac{1}{2} + \frac{1}{2}|\langle\psi|\phi\rangle|^2$$

Where if the two states in comparison are orthogonal to each other, $|\langle\psi|\phi\rangle|^2 = 0$. If the two states in comparison are equal, then $|\langle\psi|\phi\rangle|^2 = 1$. Once the ancilla qubit is measured and this fidelity value has been outputted, we determine a classification result by using a simple threshold value. If the fidelity is higher than the threshold, then the result is a face image. If the fidelity is lower than the threshold, the result is a non-face image.

### IV. RESULTS

#### A. Quantum face identification

To demonstrate the initial potential of the systems' classification ability, there has to be a substantial difference in fidelities of face and non-face images to be able to classify successfully. Various scales of input vector and qubits were compared to determine their effects on the output. For the purposes of this work, the quantum system explored was developed using the quantum software library PennyLane [29] and simulated using IBM's high-performance quantum simulators [30].

The table below shows results of the average fidelities of face and non-face images in the dataset using differing input vector dimensions and circuit sizes. We stop at 17 qubits as we feel a much more complex system size is beyond the scope of what this work is trying to present.

TABLE 1. AVERAGE FIDELITIES FOR FACE AND NON-FACE IMAGES PER INPUT SIZE

| No. of Qubits | Input Vector Dimensions | Average Face Fidelity | Average Non-Face Fidelity |
|---|---|---|---|
| 9 | 16 | 0.95434 | 0.87156 |
| 13 | 64 | 0.96199 | 0.82020 |
| 17 | 256 | 0.86848 | 0.78061 |

Here we can see that the average fidelities suggest there is a balance between system size and output. Perhaps by scaling down the input vectors, we are losing too much important detail for classification, whereas a larger system has too much variability from larger numbers of qubits, causing the fidelity to drop.

To determine the classification performance of the system, various threshold levels were compared using the test dataset consisting of 100 face and 300 non-face images. The thresholds compared started at a value of 0.7 and were incremented in steps of 0.01 until a final value of 1.0 was reached. The table and figure below display results of these tests. The results in table 2 also suggest that it is important to be able to balance the size of the system we are using to keep important detail in our input vectors whilst not using a too large system size that could be causing uncertainties in our output.

TABLE 2. ACCURACY AND BEST PERFORMING THRESHOLD VALUE PER SYSTEM SIZE

| No. of Qubits | System Accuracy | Highest Performing Threshold |
|---|---|---|
| 9 | 0.800 | 0.971 |
| 13 | 0.906 | 0.958 |
| 17 | 0.800 | 0.993 |

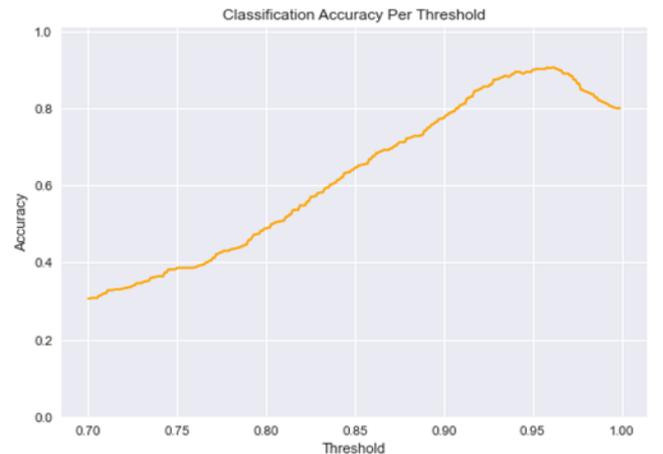

Figure 2. Graph of classification accuracy per threshold (13 qubit system)

### B. Performance comparison against classical algorithms

To demonstrate the quantum systems' viability, the previous results were compared against the common classical algorithms of a Support Vector Machine and k-Nearest Neighbor algorithms. For the purposes of these experiments with classical methods, 300 face/non-face images were used as the training portion and the remaining 100 images per class were used as the testing portion. The SVM tested used 115 support vectors and a Gaussian kernel. For the k-NN, various values of k ranging from 1 to 20 were also compared. The table and figure below show the performance of these classical algorithms compared against the proposed quantum method and the results of various k values for the k-NN comparison.

TABLE 3. ACCURACY COMPARISON BETWEEN PROPOSED SYSTEM AND CLASSICAL ALGORITHMS

| Algorithm | System Accuracy |
|---|---|
| SVM | 0.995 |
| k-NN | 0.945 (k=2) |
| Quantum | 0.906 |

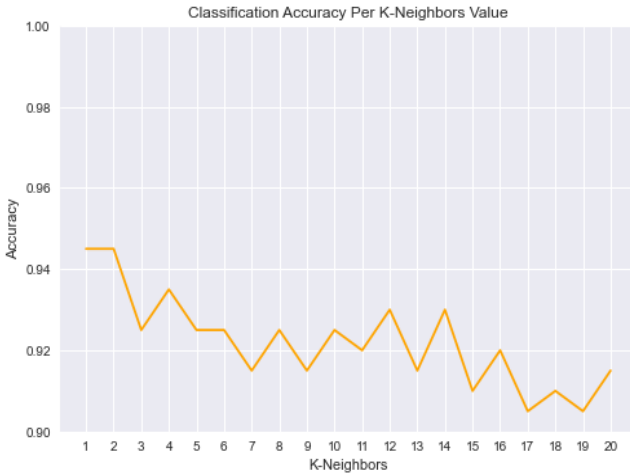

Figure 3. Classification accuracy comparison using differing k-neighbor values

Here we can see that the classical algorithms do perform to a higher level of accuracy than the quantum method explored in this work. However, the quantum system does still produce a reasonably high level of performance that could still be very feasible to use in many applications with improvements to current quantum hardware are made in time.

### C. Complexity analysis

Whilst the accuracy of these systems has been compared previously, a key aspect that is driving for the development of quantum algorithms is the speedup benefit of potentially exponential proportions over current classical methods. In order to display a fair comparison of this quantum method to classical algorithms, the computational complexity of the algorithms used must also be compared alongside the performance metric of accuracy. This can be seen in the table below.

TABLE 4. COMPUTATIONAL COMPLEXITY COMPARISON OF ALGORITHMS

| Algorithm | Complexity |
|---|---|
| SVM | $O((I^2 D) + I^3)$ |
| k-NN | $O(DI)$ |
| Quantum | $O(I \log_2(D))$ |

Where $D$ is the number of dimensions in each input vector, $N_{SV}$ is the number of support vectors used and $I$ is the number of images used. Here we can see that the proposed quantum method achieves exponential speedups in areas over the classical methods explored in this section, due to the nature of efficient quantum encoding and representation of quantum information.

## V. DISCUSSION & FUTURE WORK

From the results outlined previously, the quantum algorithm explored in this work performs at a slightly lower level when comparing the metric of accuracy in comparison to popular classical algorithms of an SVM and k-NN. However, once the computational complexity is considered in a broader scope of comparison, we can see that the speedup potential of quantum algorithms is realized.

The results outlined in section 4 suggest that we must balance the way in which we design our algorithms and find a middle ground that contains high levels of detail in our input vectors, without causing the resulting number of qubits to scale to a large size that becomes uncontrollable and limits our output potential.

The algorithm explored in this work is not without its' limitations however. More studies should be conducted into the robustness of the algorithm and how its' performance is affected when data of a wide variety of perspectives, ratios and lighting settings are introduced. Whilst the non-face images tested were not biased and consisted of a wide variety of objects of varying ratios, objects that may appear as similar shapes or contain features that are close to that of a face could affect the robustness of the system. Other methods within the pre-processing and encoding of data could also be considered that may enhance the performance of the system with little extra computation needed.

As a whole, the results produced by the quantum method explored are promising when compared to classical methods, but only once the complexities of the compared algorithms are considered. Over time as improvements to quantum hardware are being made, we may see methods such as these become much more feasible to use within real-world application settings.

## VI. CONCLUSION

In this paper we explored the concept towards the use of quantum computing for the widely-used machine learning application of facial identification. By encoding feature high-dimensional feature vectors into quantum states, complex feature spaces can be created within Hilbert space. By using the SWAP test quantum subroutine, we are able to measure the fidelity of two quantum states which can allow us to classify an image as a face or non-face objects appropriately.

Initial tests of the proposed quantum system show promising early results that can only hope to improve as more advanced quantum technologies and algorithms are presented. Whilst there are classical algorithms that are capable of outperforming the quantum system in terms of accuracy, we gain exponential speedups over these algorithms for small decreases in accuracy. In the NISQ era of quantum computing, the tradeoff of this can be considered fair and it is hoped that, in time, the quantum hardware available will be able to improve upon these benchmarks.

### CONFLICT OF INTEREST

The authors declare no conflict of interest.

### AUTHOR CONTRIBUTIONS

P. Easom-McCaldin conducted the initial investigation, prepared the data and experiments, analyzed the results and developed the manuscript. Corresponding authors A. Bouridane and A. Belatreche provided valuable insight. All authors approved the final version.

### REFERENCES

[1] S. Lloyd, M. Mohseni, and P. Rebentrost, "Quantum principal component analysis," Nat. Phys., vol. 10, no. 9, pp. 631–633, Jul. 2014, doi: 10.1038/NPHYS3029.

[2] S. Y.-C. Chen, C.-H. H. Yang, J. Qi, P.-Y. Chen, X. Ma, and H.-S. Goan, "Variational Quantum Circuits for Deep Reinforcement Learning," 2019. Accessed: May 21, 2020. [Online]. Available: http://arxiv.org/abs/1907.00397.

[3] E. Farhi and H. Neven, "Classification with Quantum Neural Networks on Near Term Processors," 2018. Accessed: May 30, 2020. [Online]. Available: http://arxiv.org/abs/1802.06002.

[4] C. M. Wilson et al., "Quantum Kitchen Sinks: An algorithm for machine learning on near-term quantum computers," 2018. Accessed: May 13, 2020. [Online]. Available: http://arxiv.org/abs/1806.08321.

[5] S. Lloyd, M. Schuld, A. Ijaz, J. Izaac, and N. Killoran, "Quantum embeddings for machine learning," 2020. Accessed: Jun. 02, 2020. [Online]. Available: http://arxiv.org/abs/2001.03622.

[6] A. Mari, T. R. Bromley, J. Izaac, M. Schuld, and N. Killoran, "Transfer learning in hybrid classical-quantum neural networks," Dec. 2019, Accessed: Apr. 28, 2020. [Online]. Available: http://arxiv.org/abs/1912.08278.

[7] I. Cong, S. Choi, and M. D. Lukin, "Quantum convolutional neural networks," 2019. doi: 10.1038/s41567-019-0648-8.

[8] M. Henderson, S. Shakya, S. Pradhan, and T. Cook, "Quanvolutional neural networks: powering image recognition with quantum circuits," 2020. doi: 10.1007/s42484-020-00012-y.

[9] N. Killoran, T. R. Bromley, J. M. Arrazola, M. Schuld, N. Quesada, and S. Lloyd, "Continuous-variable quantum neural networks," Phys. Rev. Res., vol. 1, no. 3, p. 33063, 2019, doi: 10.1103/physrevresearch.1.033063.

[10] G. R. Steinbrecher, J. P. Olson, D. Englund, and J. Carolan, "Quantum optical neural networks," npj Quantum Inf., vol. 5, no. 1, 2019, doi: 10.1038/s41534-019-0174-7.

[11] K. Beer et al., "Training deep quantum neural networks," Nat. Commun., vol. 11, no. 1, 2020, doi: 10.1038/s41467-020-14454-2.

[12] S. Garg and G. Ramakrishnan, "Advances in Quantum Deep Learning: An Overview," 2020. Accessed: May 13, 2020. [Online]. Available: http://arxiv.org/abs/2005.04316.

[13] Afham, A. Basheer, and S. K. Goyal, "Quantum k-nearest neighbor machine learning algorithm," 2020. Accessed: Jun. 22, 2020. [Online]. Available: http://arxiv.org/abs/2003.09187.

[14] Y. Ruan, X. Xue, H. Liu, J. Tan, and X. Li, "Quantum Algorithm for K-Nearest Neighbors Classification Based on the Metric of Hamming Distance," Int. J. Theor. Phys., vol. 56, no. 11, pp. 3496–3507, 2017, doi: 10.1007/s10773-017-3514-4.

[15] A. Martin et al., "Towards Pricing Financial Derivatives with an IBM Quantum Computer," 2019. Accessed: Jun. 22, 2020. [Online]. Available: http://arxiv.org/abs/1904.05803.

[16] J. M. Arrazola, A. Delgado, B. R. Bardhan, and S. Lloyd, "Quantum-inspired algorithms in practice," 2019. Accessed: Jun. 07, 2020. [Online]. Available: http://arxiv.org/abs/1905.10415.

[17] D. Kopczyk, "Quantum machine learning for data scientists," 2018. Accessed: Jun. 11, 2020. [Online]. Available: http://arxiv.org/abs/1804.10068.

[18] A. Pérez-Salinas, A. Cervera-Lierta, E. Gil-Fuster, and J. I. Latorre, "Data re-uploading for a universal quantum classifier," Quantum, vol. 4, p. 226, Jul. 2020, doi: 10.22331/q-2020-02-06-226.

[19] V. Havlíček et al., "Supervised learning with quantum-enhanced feature spaces," 2019. doi: 10.1038/s41586-019-0980-2.

[20] P. Rebentrost, M. Mohseni, and S. Lloyd, "Quantum support vector machine for big data classification," 2014. doi: 10.1103/PhysRevLett.113.130503.

[21] N. Liu and P. Rebentrost, "Quantum machine learning for quantum anomaly detection," 2018. doi: 10.1103/PhysRevA.97.042315.

[22] I. Cong and L. Duan, "Quantum discriminant analysis for dimensionality reduction and classification," New J. Phys., vol. 18, no. 7, 2016, doi: 10.1088/1367-2630/18/7/073011.

[23] C. Ding, T.-Y. Bao, and H.-L. Huang, "Quantum-Inspired Support Vector Machine," 2019. Accessed: Jun. 01, 2020. [Online]. Available: http://arxiv.org/abs/1906.08902.

[24] M. Ostaszewski, P. Gawron, and P. Sadowski, "Quantum image classification using principal component analysis," 2015. doi: 10.20904/271001.

[25] E. Şahín and I. Yilmaz, "QRMW: Quantum representation of multi wavelength images," Turkish J. Electr. Eng. Comput. Sci., vol. 26, no. 2, pp. 768–779, 2018, doi: 10.3906/elk-1705-396.

[26] [1] Y. Zhang, K. Lu, Y. Gao, and M. Wang, "NEQR: A novel enhanced quantum representation of digital images," Quantum Inf. Process., vol. 12, no. 8, pp. 2833–2860, 2013, doi: 10.1007/s11128-013-0567-z.

[27] H. S. Li, X. Chen, H. Xia, Y. Liang, and Z. Zhou, "A Quantum Image Representation Based on Bitplanes," IEEE Access, vol. 6, pp. 62396–62404, 2018, doi: 10.1109/ACCESS.2018.2871691.

[28] Le · P. Q., F. Dong, and K. Hirota, "A flexible representation of quantum images for polynomial preparation, image compression, and processing operations," vol. 10, pp. 63–84, 2011, doi: 10.1007/s11128-010-0177-y.

[29] V. Bergholm et al., "PennyLane: Automatic differentiation of hybrid quantum-classical computations." Accessed: Jul. 21, 2020. [Online]. Available: https://pennylane.ai.

[30] H. Abraham et al. "Qiskit: An Open-source Framework for Quantum Computing." 2019, doi: 10.5281/zenodo/256211